\documentclass[3p,twocolumn]{elsarticle}

\usepackage{color,soul} 
\usepackage[colorlinks=true,citecolor=blue,linkcolor=red,linktocpage=true,pagebackref=false]{hyperref}

\usepackage{amssymb}

\begin{document}




\title{Nanomagnet coupled to quantum spin Hall edge: An adiabatic
quantum motor}


\author{Liliana Arrachea}
\address{Departamento de F\'{\i}sica, FCEyN, Universidad de Buenos Aires and IFIBA, Pabell\'on I, Ciudad Universitaria, 1428 CABA  and International Center for Advanced Studies, UNSAM, Campus Miguelete, 25 de Mayo y Francia, 1650 Buenos Aires, Argentina}

\author{Felix von Oppen}
\address{Dahlem Center for Complex Quantum Systems and Fachbereich Physik, Freie Universit\"at Berlin, 14195 Berlin, Germany}

\begin{abstract}
The precessing magnetization of a magnetic islands coupled to a quantum spin Hall edge pumps charge along the edge. Conversely, a bias voltage applied to the edge makes the magnetization precess. We point out that this device realizes an adiabatic quantum motor and discuss the efficiency of its operation based on a scattering matrix approach akin to Landauer-B\"uttiker theory. Scattering theory provides a microscopic derivation of the Landau-Lifshitz-Gilbert equation for the magnetization dynamics of the device, including spin-transfer torque, Gilbert damping, and Langevin torque. We find that the device can be viewed as a Thouless motor, attaining unit efficiency when the chemical potential of the edge states falls into the magnetization-induced gap. For more general parameters, we characterize the device by means of a figure of merit analogous to the ZT value in thermoelectrics. 
\end{abstract}

\maketitle


\section{Introduction}
\label{intro}

\begin{figure}[t]
\centering
\includegraphics[width=0.47\textwidth]{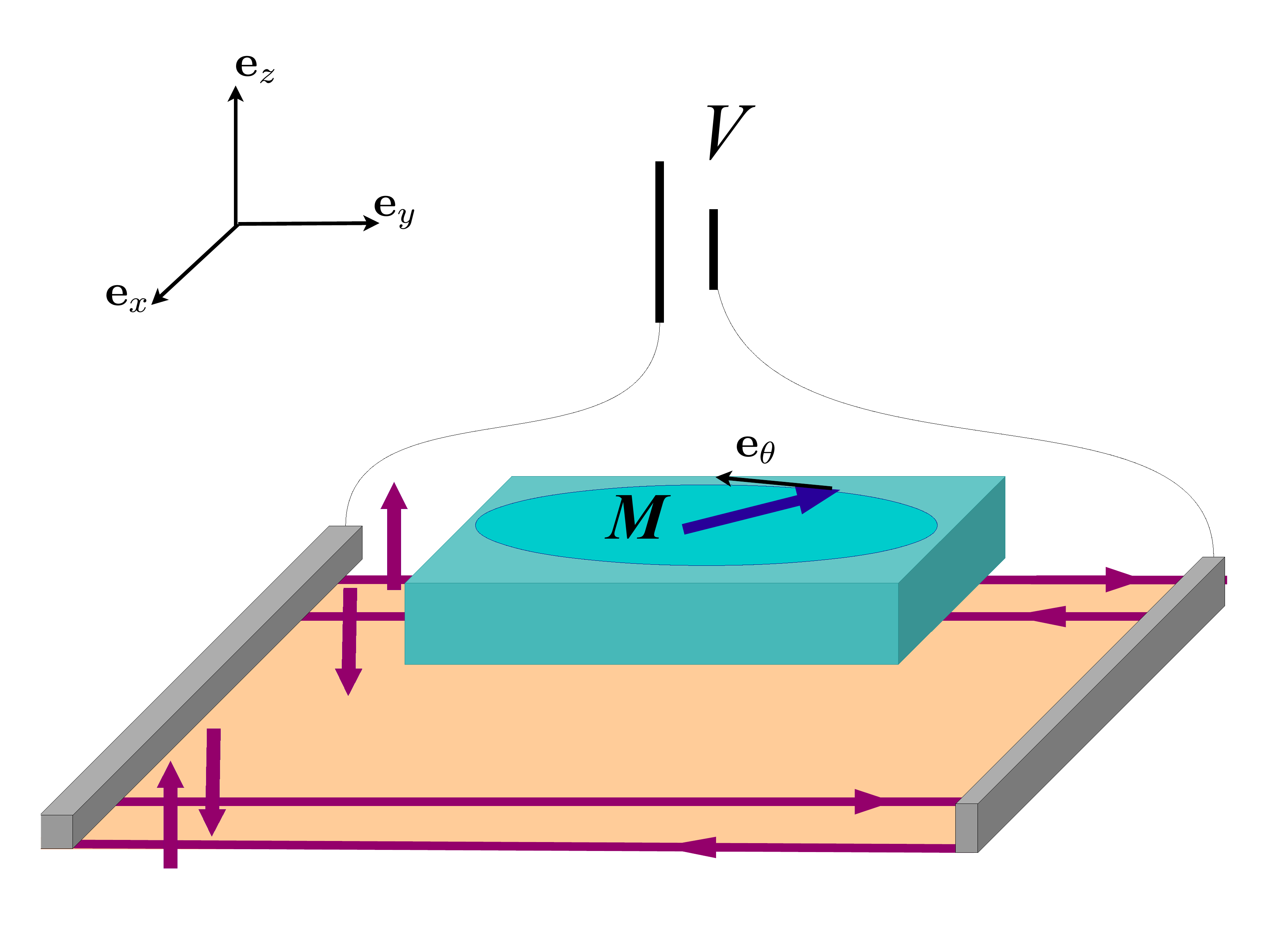}
\caption{(Color online) Schematic setup. A nanomagnet with magnetic moment ${\bf M}$ couples to a Kramers pair of edge states of a quantum spin Hall insulator.  The effective spin current produces a spin-transfer toque and the magnetic moment precesses.    }
\label{fig1}
\end{figure}

Following Ref.\ \cite{qi-zhang}, Meng {\em et al.} \cite{meng} recently showed that a transport current flowing along a quantum spin Hall edge causes a precession of the magnetization of a magnetic island which locally gaps out the edge modes (see Fig.\ \ref{fig1} for a sketch of the device). The magnetization dynamics is driven by the spin transfer torque exerted on the magnetic island by electrons backscattering from the gapped region. Indeed, the helical nature of the edge state implies that the backscattering electrons reverse their spin polarization, with the change in angular momentum transfered to the magnetic island. This effect is not only interesting in its own right, but may also have applications in spintronics. 

Current-driven directed motion at the nanoscale has also been studied for mechanical degrees of freedom, as motivated by progress on nanoelectromechanical systems. Qi and Zhang \cite{qi2009} proposed that a conducting helical molecule placed in a homogeneous electrical field could be made to rotate around its axis by a transport current and pointed out the intimate relations with the concept of a Thouless pump \cite{Thouless}. Bustos-Marun {\em et al.} \cite{bus} developed a general theory of such adiabatic quantum motors, used it to discuss their efficiency, and emphasized that the Thouless motor discussed by Qi and Zhang is optimally efficient. 

It is the purpose of the present paper to emphasize that the current-driven magnetization dynamics is another -- perhaps more experimentally feasible -- variant of a Thouless motor and that the theory previously developed for adiabatic quantum motors \cite{bus} is readily extended to this device. This theory not only provides a microscopic derivation of the Landau-Lifshitz-Gilbert equation for the current-driven magnetization dynamics, but also allows one to discuss the efficiency of the device and to make the relation with the magnetization-driven quantum pumping of charge more explicit. 

Specifically, we will employ an extension of the Landauer-B\"uttiker theory of quantum transport which includes the forces exerted by the electrons on a slow classical degree of freedom \cite{niels1,niels1p,thom,niels2}. Markus B\"uttiker developed Landauer's vision of quantum coherent transport as a scattering problem into a theoretical framework \cite{scat,noise1} and applied this scattering theory of quantum transport to an impressive variety of phenomena. These applications include Aharonov-Bohm oscillations \cite{ab}, shot noise and current correlations \cite{noise1,noise2,noise3}, as well as edge-state transport in the integer Hall effect \cite{qhall} and topological insulators \cite{topin}. Frequently, B\"uttiker's predictions based on scattering theory provided reference points with which other theories -- such as the Keldysh Green-function formalism \cite{past,jau,pump3,trans} or  master equations \cite{han} -- sought to make contact. 
  
In the present context, it is essential that scattering theory also provides a natural framework to study quantum coherent transport in systems under time-dependent driving. For adiabatic driving, B\"uttiker's work with Thomas and Pr\^etre \cite{BTP} was instrumental in developing a description of adiabatic quantum pumping \cite{Thouless} in terms of scattering theory \cite{brou,pump1,pump2,Moskalets04} which provided a useful backdrop for later experiments \cite{switkes,saw,expump2,expump3,expump4}. Beyond the adiabatic regime, Moskalets and B\"uttiker combined the scattering approach with Floquet theory to account for periodic driving \cite{Moskalets2002}. These works describe adiabatic quantum transport as a limit of the more general problem of periodic driving and ultimately triggered numerous studies on single-particle emitters and quantum capacitors (as reviewed by Moskalets and Haack in this volume \cite{haack}). 

The basic idea of the adiabatic quantum motor \cite{bus} is easily introduced by analogy with the Archimedes screw, a device consisting of a screw inside a pipe. By turning the screw, water can be pumped against gravity. This is a classical analog of a quantum pump in which electrons are pumped between reservoirs by applying periodic potentials to a central scattering region. Just as the Archimedes pump can pump water against gravity, charge can be quantum pumped against a voltage. In addition, the Archimedes screw has an inverse mode of operation as a {\em motor}: Water pushed through the device will cause the screw to rotate. The adiabatic quantum motor is a quantum analog of this mode of operation in which a transport current pushed through a quantum coherent conductor induces unidirectional motion of a classical degree of freedom such as the rotations of a helical molecule. 

The theory of adiabatic quantum motors \cite{bus,marun} exploits the assumption that the motor degrees of freedom -- be they mechanical or magnetic -- are slow compared to the electronic degrees of freedom. In this adiabatic regime, the typical time scale of the mechanical dynamics is large compared to the dwell time of the electrons in the interaction region between motor and electrical degrees of freedom. In this limit, the dynamics of the two degrees of freedom can be discussed in a mixed quantum-classical description. The motor dynamics is described in terms of a classical equation of motion, while a fully quantum-coherent description is required for the fast electronic degrees of freedom. 

From the point of view of the electrons, the motor degrees of freedom act as {\em ac} potentials which pump charge through the conductor. Conversely, the backaction of the electronic degrees of freedom enters through adiabatic reaction forces on the motor degrees of freedom \cite{niels1,niels1p,thom,niels2}. When there is just a single (Cartesian) classical degree of freedom, these reaction forces are necessarily conservative, akin to the Born-Oppenheimer force in molecular physics \cite{pisto}. Motor action driven by transport currents can occur when there is more than one motor degree of freedom (or a single angle degree of freedom). In this case, the adiabatic reaction force need no longer be conservative when the electronic conductor is subject to a bias voltage \cite{niels1,niels1p,thom,niels2}. 

In next order in the adiabatic approximation, the electronic system also induces frictional and Lorentz-like forces, both of which are linear in the slow velocity of the motor degree of freedom. Including the fluctuating Langevin force which accompanies friction yields a classical Langevin equation for the motor degree of freedom. This equation can be derived systematically within the Keldysh formalism \cite{pisto} and the adiabatic reaction forces expressed through the scattering matrix of the coherent conductor \cite{niels1,niels1p,thom}.

While these developments focused on mechanical degrees of freedom, it was also pointed out that the scattering theory of adiabatic reaction forces extends to magnetic degrees of freedom \cite{niels2}. In this case, adiabaticity requires that the precessional time scale of the magnetic moment is larger than the electronic dwell time. The effective classical description for the magnetic moment takes the form of a Landau-Lifshitz-Gilbert (LLG) equation. Similar to nanoelectromechanical systems, the LLG equation can be derived systematically in the adiabatic limit for a given microscopic model and the coefficients entering the LLG equation can be expressed alternatively in terms of electronic Green functions or scattering matrices  \cite{brat1,brat2,brat3,brat4,niels2}. In the following, we will apply this general theory to a magnetic island coupled to a Kramers pair of helical edge states.  

This work is organized as follows. Section 2 reviews the scattering-matrix expressions for the torques entering the LLG equation. Section 3 applies this theory to helical edge states coupled to a magnetic island and makes the relation to adiabatic quantum motors explicit. Section 4 defines and discusses the efficiency of this device and derives a direct relation between charge pumping and spin transfer torque. Section 5 is devoted to conclusions.  

\section{S-matrix theory of spin transfer torques and Gilbert damping}
\label{smat}

\subsection{Landau-Lifshitz-Gilbert equation}
Consider a coherent (Landauer-B\"uttiker) conductor coupled to a magnetic moment. The latter is assumed to be sufficiently large to justify a classical description of its dynamics but sufficiently small so that we can treat it as a single macrospin. Then,  its dynamics is ruled by a Landau-Lifshitz-Gilbert equation 
\begin{equation} 
\label{llg1}
  \dot{\bf M} = {\bf M} \times \left[ - \partial_{\bf M} U + {\bf B}_{\rm el} + \delta {\bf B} \right].
\end{equation}
Note that we use units in which ${\bf M}$ is an angular momentum and for simplicity of notation, ${\bf B}_{\rm el}$ as well as $\delta {\bf B}$ differ from a conventional magnetic field by a factor of $g_d$, the gyromagnetic ratio of the macrospin. The first term on the right-hand side describes the dynamics of the macrospin in the absence of coupling to the electrons. It is derived from the quantum Hamiltonian
\begin{equation}
 \hat{U}= - g_d \hat{\bf M} \cdot {\bf B} + \frac{D}{2} \hat{M}_z^2,
\end{equation}
where ${\bf M}=\langle \hat{\bf M} \rangle$ is the uncoupled macrospin, ${\bf B}$ the magnetic field, and $D>0$ the easy-plane anisotropy of the macrospin. The coupling to the electrons leads to the additional effective magnetic field ${\bf B}_{\rm el}$. This term can be derived microscopically from the Heisenberg equation of motion of the macrospin by evaluating the commutator of $\hat{\bf M}$ with the interaction Hamiltonian between macrospin and electrons in the adiabatic approximation (see, e.g., Ref.\ \cite{niels2}). Keeping terms up to linear order in the small magnetization ``velocity'' $\dot {\bf M}$, we can write
\begin{equation}
   {\bf B}_{\rm el} = {\bf B}_{0}({\bf M}) - \gamma({\bf M}) \dot{\bf M}.
\label{Bel}
\end{equation}
Here, the first contribution ${\bf B}_0$ can be viewed as the spin-transfer torque. The second term is a contribution to Gilbert damping arising from the coupling between macrospin and electrons. In general, $\gamma$ so derived is a tensor with symmetric and antisymmetric components. However, it can be seen that only the symmetric part plays a relevant role \cite{niels2}. Finally, by fluctuation-dissipation arguments, the Gilbert damping term is accompanied by a Langevin torque $\delta {\bf B}$ with correlator
\begin{equation}
 \langle \delta { B}_l (t)\delta { B}_k (t^{\prime} ) \rangle = {\cal D}_{lk} \delta (t-t^{\prime}).
\end{equation}
Its correlations are local in time as a consequence of the assumption of adiabaticity. As a result, we find the LLG equation
\begin{equation} 
\label{llg}
  \dot{\bf M} = {\bf M} \times \left[ - \partial_{\bf M} U +  {\bf B}_0 - \gamma  \dot{\bf M}  + \delta {\bf B} \right],
\end{equation}
for the macrospin ${\bf M}$.

The spin-transfer torque, the Gilbert damping, and the correlator ${\cal D}$ can be expressed in terms of the scattering matrix of the coherent conductor, both in and out of equilibrium \cite{brat1,brat2,brat3,brat4,niels2}. Before presenting the S-matrix expressions, a few comments are in order. First, the expression for the Gilbert damping only contains the intrinsic damping originating from the coupling to the electronic degrees of freedom. Coupling to other degrees of freedom might give further contributions to Gilbert damping which could be included phenomenologically. Second, in the study of the nanomagnet coupled to the helical modes we will consider the expressions to lowest order in the adiabatic approximation presented in Sec.\ \ref{lowest}. The theory can actually be extended to include higher order corrections \cite{niels2}.  In Sec.\ \ref{higher} section, we briefly summarize the main steps of the general procedure for completeness.

\subsection{Coefficients of the LLG equation in the lowest order adiabatic approximation}\label{coef}
\label{lowest}

This section summarizes the expressions for the coefficients of the LLG equation that we will use to study the problem of the nanomagnet coupled to the helical edge states. These correspond to the lowest order in the adiabatic approximation, in which we retain only terms linear in $\dot{\bf M}$ and $eV$. 

To this order, we can write the coefficients of the LLG equation in terms of the electronic S-matrix for a static macrospin ${\bf M}$. The coupling between macrospin and electronic degrees of freedom enters through the dependence of the electronic S-matrix $S_0 = S_0({\bf M})$ on the (fixed) macrospin. At this order, the spin-transfer torque and the Gilbert damping can be expressed as \cite{brat1,brat2,brat3,brat4,niels2}
\begin{equation}
 {\bf B}_0({\bf M})=  \sum_{\alpha}  \int \frac{d \varepsilon}{2 \pi i} f_{\alpha} \mbox{Tr}\left[ \Pi_{\alpha} \hat{S}_0^{\dagger} \frac{ \partial       S_0}{\partial {\bf M}} \right]
 \label{STTS}
\end{equation}
and 
\begin{eqnarray} \label{gil1}
 \gamma^{kl}({\bf M})  =  -  \hbar \sum_{\alpha} \int \frac{d \varepsilon}{ 4 \pi}  f^{\prime}_{\alpha}
   \mbox{Tr}\left[ \Pi_{\alpha} \frac{ \partial \hat{S}^{\dagger}_0}{\partial M_k}  \frac{ \partial \hat{S}_0}{\partial M_l} \right]_s, 
\end{eqnarray}
respectively. 
 Finally, the fluctuation correlator ${\cal D}$ is expressed as \cite{niels2}
\begin{eqnarray} 
  \label{fluc}
 & &  {\cal D}^{kl}({\bf M}) = \hbar \sum_{\alpha, \alpha^{\prime}} \int \frac{ d\varepsilon}{2 \pi} f_{\alpha} \left( 1- f_{\alpha^{\prime}} \right)     
       \nonumber \\
    && \times\mbox{Tr}\left[ \Pi_{\alpha} \left( \hat{S}^{\dagger}_0 \frac{\partial \hat{S}_0}{\partial M_k} \right)^{\dagger} \Pi_{\alpha^{\prime}}    
       \left(\hat{S}^{\dagger}_0 \frac{\partial \hat{S}_0}{\partial M_l} \right) \right]_s . 
\end{eqnarray}
In these expressions, $\alpha=L,R$ denotes the reservoirs with electron distribution function $f_\alpha$, $\Pi_{\alpha}$ is a projector onto the channels of lead $\alpha$, and $\mbox{Tr}$ traces over the lead channels.

\subsection{Corrections to the adiabatic approximation of the S-matrix}
\label{higher}

In order to go beyond linear response in $eV$ and $\dot{\bf M}$, we must consider the electronic S-matrix in the presence of the {\em time-dependent} magnetization ${\bf M}(t)$ and expand it to linear order in the magnetization ``velocity'' $\dot {\bf M}(t)$. This can be done, e.g., by starting from the full Floquet scattering matrix $S^F_{\alpha, \beta}(\varepsilon_n,\varepsilon)$ for a periodic driving with period $\omega$ \cite{Moskalets2002}. The indices $\alpha$ and $\beta$ label the scattering channels of the coherent conductor and the arguments denote the energies $\varepsilon$ of the incoming electron in channel $\alpha$ and $\varepsilon_n=\varepsilon + n  \hbar \omega$ of the outgoing electron in channel $\beta$. For small driving frequency $\omega$, the Floquet scattering matrix can be expanded in powers of $\hbar \omega$, 
\begin{eqnarray}
   \hat{S}^F (\varepsilon_n,\varepsilon)&=& \hat{S}^0_n (\varepsilon) + \frac{n \hbar \omega}{2} \frac{\partial  \hat{S}^0_n (\varepsilon) }{\partial 
     \varepsilon} \nonumber \\
     & & + \;  \hbar \omega \hat{A}_n(\varepsilon) + {\cal O}(\varepsilon^2).
\end{eqnarray}
Here $ \hat{S}^0_n (\varepsilon) $ is the Fourier transform of the frozen scattering matrix $S_0({\bf M}(t))$ introduced above, 
\begin{equation}
    \hat{S}_0({\bf M}(t))= \sum_{n=-\infty}^{\infty} e^{-i n \omega t}\hat{S}^0_n (\varepsilon).
\end{equation}
The matrix $\hat{A}_n(\varepsilon)$, first introduced by Moskalets and B\"uttiker, is the first adiabatic correction to the adiabatic S-matrix and 
can be transformed in a similar way to 
\begin{equation}
 \hat{\cal A}(t, \varepsilon)= \sum_{n=-\infty}^{\infty} e^{-i n \omega t}\hat{A}_n (\varepsilon)= \dot{\bf M}(t) \cdot \hat{\bf A}(t, \varepsilon) .
\end{equation}
The matrix $\hat{A}_n(\varepsilon)$ can be straightforwardly calculated from the retarded Green function of the device (see Refs. \cite{trans,niels2}).  

We are now in a position to give expressions for the Gilbert damping to next order in the adiabatic approximation. (The spin-transfer torque and the fluctuation correlator remain unchanged.) To do so, we split the Gilbert matrix $\gamma$ into its symmetric and antisymmetric parts,
\begin{equation}
  \gamma = \gamma_s + \gamma_a.
\end{equation}
Strictly speaking, it is only the symmetric part which corresponds to Gilbert damping. The antisymmetric part simply renormalizes the precession frequency. One finds  \cite{niels2}
\begin{eqnarray} \label{gil}
   \gamma_s^{kl}({\bf M})  =  
 -  \hbar \sum_{\alpha} \int \frac{d \varepsilon}{ 4 \pi}  f^{\prime}_{\alpha}
 \mbox{Tr}\left[ \Pi_{\alpha} 
 \frac{ \partial \hat{S}^{\dagger}_0}{\partial M_k}  \frac{ \partial \hat{S}_0}{\partial M_l} \right]_s & &  \nonumber \\
  + \sum_{\alpha} \int \frac{d \varepsilon}{2 \pi i } f_{\alpha}
 \mbox{Tr}\left[ \Pi_{\alpha} \left( \frac{\partial \hat{S}^{\dagger}_0}{\partial M_k}  \hat{A}_l
 - \hat{A}_l^{\dagger} \frac{\partial \hat{S}_0}{\partial M_k} \right) \right]_s & & 
\end{eqnarray}
for the symmetric contribution. It can be seen, the second line is a pure nonequilibrium contribution ($\propto eV \hbar \omega$). Similarly, the antisymmetric part of the Gilbert damping can be written as \cite{niels2} 
\begin{eqnarray}
 &  & \gamma_a^{kl}({\bf M})   =  
 - \hbar \sum_{\alpha} \int \frac{d \varepsilon}{ 2 \pi i }  f_{\alpha}(\varepsilon) \nonumber  \\
&  & \times \mbox{Tr}\left[ \Pi_{\alpha} 
  \left( \hat{S}^{\dagger}_0 \frac{\partial\hat{A}_k  }{\partial M_l}  
 - \frac{\partial  \hat{A}_k^{\dagger}}{\partial M_l}  \hat{S}_0 \right) \right]_a,
\end{eqnarray}
which is really a renormalization of the precession frequency as mentioned above.

\section{S-matrix theory of a nanomagnet coupled to a quantum spin Hall edge}

We now apply the above theory to a magnetic island coupled to a quantum spin Hall edge as sketched in Fig.\ \ref{fig1}. The quantum spin Hall edge supports a Kramers doublet of edge states. The magnetization ${\bf M} = M_\perp \cos \theta {\bf e}_x + M_\perp \sin \theta {\bf e}_y + M_z {\bf e}_z$ of the magnetic island induces a Zeeman field $J{\bf M}$ acting on the electrons along the section of length $L$ of the edge state which is covered by the magnet. This Zeeman field causes backscattering between the edge modes and induces a gap $\Delta=J M_\perp \hbar/2$ \cite{qi-zhang}. Linearizing the dispersion of the edge modes, the electronic Hamiltonian takes the form \cite{meng} 
\begin{equation}
\hat{H} =  (v p -JM_z)\hat{\sigma}_z  +  \Delta(x) \left( \cos \theta \hat{\sigma}_x + \sin \theta \hat{\sigma}_y \right).
\label{elH}
\end{equation}
Here, the $\sigma_j$ denote Pauli matrices in spin space and $\Delta(x)$ is nonzero only over the region of length $L$ covered by the magnetic island. We have assumed for simplicity that the spin Hall edge conserves $\sigma_z$. Then, a static island magnetization induces a gap whenever it has a component perpendicular to the $z$-direction. Indeed, $\hat H$ is easily diagonalized for a spatially uniform coupling between edge modes and magnet, and the spectrum 
\begin{equation}
  E_p =  \sqrt{ ( v p - J M_z)^2 + \Delta^2}
\end{equation}
has a gap $\Delta$.

In the following, we assume that the easy-plane anisotropy $D>0$ is sufficiently large so that the magnetization entering the electronic Hamiltonian can be taken in the $xy$-plane, i.e., $M_z\simeq 0$. (However, we will have to keep $M_z$ in the LLG equation when it is multiplied by the large anisotropy $D$.) 

The electronic Hamiltonian (\ref{elH}) is equivalent to the electronic Hamiltonian of the Thouless motor considered in Ref.\ \cite{bus}. Following this reference, we can readily derive the frozen scattering matrix analytically \cite{bus},
\begin{equation}\label{s-matrix0}
  \hat{S}_0 = \frac{1}{\Lambda} \left( \begin{array}{cc} -i e^{i \theta} \lambda & 1 \\
       1 &  -i e^{-i \theta} \lambda  \end{array} \right) ,  
\end{equation}
where we have defined the shorthands
\begin{eqnarray}
   \Lambda&=& \cos \phi_L - i \frac{\varepsilon}{\sqrt{\varepsilon^2 - \Delta^2}} \sin \phi_L, \nonumber \\
   \lambda &=&  \frac{\Delta}{\sqrt{\varepsilon^2 - \Delta^2}} \sin \phi_L
\end{eqnarray} 
with
\begin{equation}
 \phi_L(\varepsilon)=\frac{L}{\hbar v} \sqrt{\varepsilon^2 - \Delta^2}.
\end{equation}
Note that these expressions are exact for any $L$ and valid for energies $\varepsilon$ both inside and outside the gap. 

We can now use this scattering matrix to evaluate the various coefficients in the LLG equation, employing the expressions given in Sec. \ref{coef}. Assuming zero temperature, we find
\begin{equation}\label{stt}
   {\bf B}_0 =  \frac{eV}{2 \pi M} \xi(\mu) {\bf e}_{\theta},
\label{B0STT}
\end{equation}
for the spin transfer torque at arbitrary chemical potential $\mu$. Here, we have defined the function
\begin{equation} \label{xi}
 \xi(\mu)= \frac{  {\Delta}^2\sin^2 \phi_L}{ |\mu^2- {\Delta}^2| \cos^2 \phi_L+ \mu^2 \sin^2 \phi_L}
\end{equation}
with $\phi_L = \phi_L(\mu)$ (see Fig.\ \ref{fig2}). Below, we will identify $\xi$ with the charge pumped between the reservoirs during one precessional period of the magnetization ${\bf M}$. The vector ${\bf B}_0$ points in the azimuthal direction in the magnetization plane and indeed corresponds to a spin-transfer torque. Similarly, we can substitute Eq.\ (\ref{s-matrix0}) into Eq.\ (\ref{gil}) for the Gilbert damping and find that the only nonzero component of the tensor $\gamma$ is 
\begin{equation}
 \gamma_{\theta\theta} = \frac{\hbar}{2 \pi M^2} \xi(\mu).
\label{gammathetatheta}
\end{equation}
Similarly,
\begin{equation}
 {\cal D}_{\theta\theta} =  \frac{\hbar eV }{\pi M^2 } \xi(\mu) 
\end{equation}
is the only nonzero component of the fluctuation correlator. It is interesting to note that this yields an effective fluctuation-dissipation relation ${\cal D}_{\theta\theta}= 2 T_{\rm eff} \gamma_{\theta\theta}$ with effective temperature $T_{\rm eff}=eV$.

With these results, we can now write the LLG equation for the nanomagnet coupled to the helical edge state,
\begin{eqnarray}\label{llgnan}
   \dot{\bf M} & = & D {\bf M} \times M_z {\bf e}_z +   \frac{\xi (eV - \hbar \dot\theta) } {2 \pi   M}  {\bf M} \times {\bf e}_{\theta}
 \nonumber \\
 & & +  {\bf M} \times  \delta{\bf B},
\end{eqnarray}
where $\xi = \xi(\mu)$, we have expressed $\dot{\bf M} \simeq M \dot\theta {\bf e}_\theta$, and assumed zero external magnetic field ${\bf B}$. This completes our scattering-theory derivation of the LLG equation and generalizes the result obtained in Ref.\ \cite{meng} on phenomenological grounds in several respects. Equation (\ref{llgnan}) applies also for finite-length magnets and chemical potentials both inside and outside the magnetization-induced gap of the edge-state spectrum. Moreover, the identification of the $\dot\theta$-term as a damping term necessitates the inclusion of the Langevin torque $\delta {\bf B}$. Indeed, Ref.\ \cite{meng} refers to the entire term involving $eV-\hbar\dot\theta$ as the spin-transfer torque. In contrast, our derivation produces the term involving $eV$ already in zeroth order in magnetization ``velocity'' $\dot {\bf M}$, while the $\dot\theta$ term appears only to linear order. Thus, the latter term is really a conrtribution to damping and related to the energy dissipated in the electron system due to the time dependence of the magnetization. 
 
\section{Efficiency of the nanomagnet as a motor}

While the electronic Hamiltonian for the edge modes is equivalent to that of the Thouless motor discussed in Ref.\ \cite{bus}, the LLG equation for the macrospin differ from the equation of motion of the mechanical degrees of freedom discussed in Ref.\ \cite{bus}. In this section, we discuss the energetics and the efficiency of the magnetic Thouless motor against the backdrop of its mechanical cousin. 

The dynamics of the macrospin is easily obtained from the LLG equation (\ref{llgnan}) \cite{meng}. For a large anisotropy and thus small $M_z$, we need to retain the $z$-component of ${\bf M}$ only in combination with the large anisotropy $D$. Then, the steady-state value of $M_z$ is fixed by the $\theta$-component of the LLG equation,
\begin{equation}
   M_z = -\frac{ \dot\theta}{D}.
\end{equation}
The precessional motion of ${\bf M}$ about the $z$-axis is governed by the $z$-component of the LLG equation, which yields
\begin{equation}
   \dot \theta = \frac{eV}{\hbar}
\end{equation} 
and hence $M_z = -eV/(\hbar D)$. It is interesting to note that the angular frequency $\dot\theta$ of the precession is just given by the applied bias voltage, independent of the damping strength. This should be contrasted with the mechanical Thouless motor. Here, the motor degree of freedom satisfies a Newton equation of motion which is second order in time. Thus, the frequency of revolution is inversely proportional to the damping coefficient.

In steady state, the magnetic Thouless motor balances the energy provided by the voltage source through the spin-transfer torque ${\bf B}_0$ against the dissipation through Gilbert damping due to the intrinsic coupling between magnetic moment and electronic degrees of freedom. It is instructive to look at these contributions independently. The work performed by the spin-transfer torque per precessional period is given by  
\begin{equation}
    \Delta W_{\rm spin-transfer} =  \int_0^{2\pi/\dot\theta} {\mathrm d}t {\bf B}_0\cdot \dot{\mathbf M}. 
\label{deltaw}
\end{equation}
Writing this as an integral over a closed loop of the magnetization ${\bf M}$ and inserting the S-matrix expression (\ref{STTS}), we find
\begin{eqnarray}
    && \Delta W_{\rm spin-transfer}  =    \sum_{\alpha}  \int \frac{d \varepsilon}{2 \pi i} f_{\alpha} \nonumber \\
    && \,\,\,\,\,\,\,\,\,\, \times\oint {\mathrm d}{\mathbf M}\cdot  \mbox{Tr}\left[ \Pi_{\alpha} \hat{S}_0^{\dagger} \frac{ \partial \hat{S}_0}{\partial {\bf M}} \right]. 
\end{eqnarray}
Without applied bias, the integrand is just the gradient of a scalar function and the integral vanishes. Thus, we expand to linear order in the applied bias and obtain
\begin{eqnarray}
 & &    \Delta W_{\rm spin-transfer}  =   \frac{i eV}{4\pi}   \nonumber \\
    & &\times\sum_{\alpha}\oint {\mathrm d}{\mathbf M} \cdot  \mbox{Tr}\left[ (\Pi_L-\Pi_R) \hat{S}_0^{\dagger} \frac{ \partial \hat{S}_0}{\partial {\bf M}} \right].
    \label{DeltaWST} 
\end{eqnarray}
Comparing Eq.\ (\ref{DeltaWST}) with the familiar S-matrix expression for the pumped charge \cite{brou}, the right-hand side can now be identified as the bias voltage multiplied by the charge pumped between the reservoirs during one revolution of the magnetization, 
\begin{equation}
    \Delta W_{\rm spin-transfer} =  Q_p V. 
\end{equation}
With every revolution of the magnetization, a charge $Q_p$ is pumped between the reservoirs. The corresponding gain $Q_p V$ in electrical energy is driving the magnetic Thouless motor. This result can also be written as
\begin{equation}
   \dot W_{\rm spin-transfer} =  \frac{Q_pV}{2\pi}\dot\theta
   \label{PowerSpinTransfer}
\end{equation}
for the power provided per unit time by the voltage source. 

The relation between spin-transfer torque and pumped charge also allows us to identify the function $\xi(\mu)$ appearing in the LLG equation as the charge in units of $e$ pumped between the reservoirs during one precessional period of the macrospin, 
\begin{equation}
   Q_p = e \xi.
\end{equation}
This can be obtained either by deriving the pumped charge explicitly from the S-matrix expression or by evaluating Eq.\ (\ref{deltaw}) using the explicit expression Eq.\ (\ref{B0STT}).

The electrical energy gain is compensated by the energy dissipated through Gilbert damping. The dissipated energy per period is given by 
\begin{eqnarray}
  \Delta W_{\rm Gilbert} & = &   \int_0^{2\pi/\dot\theta} {\mathrm d}t \dot{\bf M}^T \gamma \dot{\bf M}^T \nonumber \\
  & = &   2\pi M^2 \gamma_{\theta\theta} \dot\theta.  
\end{eqnarray}
Using Eq.\ (\ref{gammathetatheta}), this yields the dissipated energy
\begin{equation}
  \Delta W_{\rm Gilbert} =  \xi \hbar \dot\theta
\end{equation}
per precessional period or
\begin{equation}
  \dot W_{\rm Gilbert} =  \frac{\xi \hbar}{2\pi }{\dot\theta}^2 
\end{equation}
per unit time. These expressions have a simple interpretation. Due to the finite frequency of the magnetization precession, each pumped charge absorbs on average an energy $\hbar\dot\theta$ which is then dissipated in the reservoirs. 

\begin{figure}[t]
\centering
\includegraphics[width=0.55\textwidth]{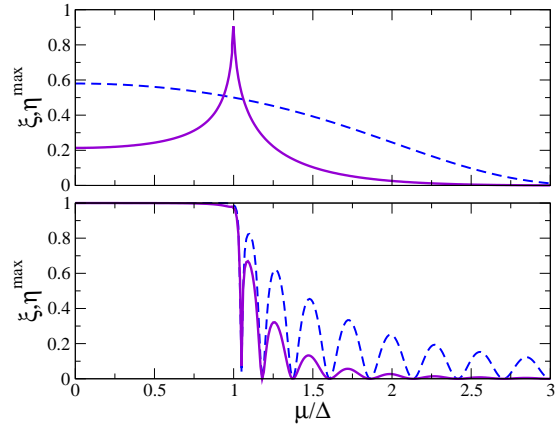}
\caption{(Color online) The parameter $\xi$ (dashed lines) entering the coefficients of the LLG equation and the maximal efficiency $\eta^{\rm max}$ (solid lines) of the motor for a fixed voltage $V$. Upper and lower panels correspond to nanomagnets of length $L= \hbar v/\Delta$ and  $L= 10 \hbar v/\Delta$, respectively.}
\label{fig2}
\end{figure}

Armed with these results, we can finally discuss the efficiency of a magnetic Thouless motor and follow the framework introduced in Ref.\ \cite{recentus} to define an appropriate figure of merit (analogous to the ZT value of thermoelectrics). Imagine the same setup as in Fig.\ \ref{fig1}, but with an additional load coupled to the magnetization. We can now define the efficiency of the magnetic Thouless motor as the ratio of the power delivered to the load and the electrical power $IV$ provided by the voltage source. In steady state, the power delivered to the load has to balance against the power provided by the electrons, i.e., ${\bf B}_{\rm el}\cdot \dot{\bf M}$. Thus, we can write the efficiency as 
\begin{equation}
    \eta = \frac{\dot W}{IV},
\end{equation}
where
\begin{eqnarray}
   \dot W & = &  \dot W_{\rm spin-torque} - \dot W_{\rm Gilbert} \nonumber \\
   & = &   \frac{\xi}{2\pi}eV\dot\theta -  \frac{\xi \hbar}{2\pi } \dot\theta^2.  
\end{eqnarray}
The total charge current  flowing along the topological insulator edge averaged over the cycle is the sum of the {\em dc} current $GV$ driven by the voltage, where $G$ is the {\em dc} conductance of the device, and the pumping current $Q_p \dot\theta/(2\pi)$,
\begin{equation}
I =  G V +  \frac{e\xi}{2 \pi} \dot\theta.
\end{equation}
We can now optimize the efficiency of the motor at a given bias $V$ as function of the frequency $\dot\theta$ of the motor revolution. Note that due to the load, the latter is no longer tied to the bias voltage $eV$. This problem is analogous to the problem of the optimal efficiency of a thermoelectric device which leads to the definition of the important ZT value. This analogy was discussed explicitly in Ref.\ \cite{recentus}. Applying the results of this paper to the present device yields the maximal efficiency 
\begin{equation}
\eta^{\rm max}=\frac{ \sqrt{1+\zeta}-1}{\sqrt{1+\zeta}+1},
\end{equation}
with a figure of merit $\zeta$ analogous to the ZT value defined by
\begin{eqnarray}
\zeta =  \frac{e^2 \xi(\mu)}{h G(\mu)},
\end{eqnarray}
where $\xi(\mu)$ is defined in Eq.\ (\ref{xi}) and the conductance reads
\begin{equation}
G(\mu)= \frac{e^2}{h} \frac{|\mu^2 - \Delta^2|^2}{|\mu^2 - \Delta^2|\cos^2 \phi_L + \mu^2 \sin^2 \phi_L}
\end{equation}
as obtained from the Landauer-B\"uttiker equation. 

As in thermoelectrics, the maximum efficiency is realized for $\zeta \rightarrow \infty$ which requires a finite pumped charge at zero conductance. Unlike thermoelectrics, the motor efficiency is bounded by $\eta=1$ instead of the Carnot efficiency. This reflects the fact that electrical energy can be fully converted into magnetic energy. Specifically, unit efficiency is reached in the limit of a true Thouless motor with zero transmission when the Fermi energy falls into the gap and nonzero and quantized pumped charge per period. This can be realized to a good approximation for a sufficiently long magnet, as seen from the lower panel in Fig.\ \ref{fig2}. For chemical potentials outside the gap, the conductance and the pumped charge exhibit  Fabry-Perot resonances. This yields a distinct sequence of maxima and minima in the efficiency. For shorter magnets, the conductance remains nonzero  within the gap, leading to lower efficiencies. This is shown in the upper panel of Fig. \ \ref{fig2}. Moreover, the Fabry-Perot resonances are washed out, so that there is only a feature at the gap edge where the conductance vanishes while $\xi \rightarrow 1/2$ for arbitrary $L$.

\section{Conclusions}

Implementing directional motion of a mechanical or magnetic degree of freedom is a fundamental problem of nanoscale systems. An attractive general mechanism relies on running quantum pumps in reverse. This is the underlying principle of adiabatic quantum motors which drive periodic motion of a classical motor degree of freedom by applying a transport current. In this paper, we emphasize that a magnetic island coupled to a quantum spin Hall edge, recently discussed by Meng {\em et al.} \cite{meng}, is just such an adiabatic quantum motor. We derive the Landau-Lifshitz-Gilbert equation for the magnetization dynamics from a general scattering-theory approach to adiabatic quantum motors, providing a microscopic derivation of spin-transfer torque, Gilbert damping, and Langevin torque. This approach does not only provide a detailed microscopic understanding of the operation of the device but also allows one to discuss its efficiency. We find that the device naturally approaches optimal efficiency when the chemical potential falls into the magnetization-induced gap and the conductance is exponentially suppressed. This makes this system a Thouless motor and possibly its most experimentally feasible variant to date. 

Several issues are left for future work. While we derived microscopic expressions for the Langevin torque, we have not explored its consequences for the motor dynamics. It should also be interesting to consider thermal analogs driven by a temperature gradient instead of a bias voltage. Inducing the magnetization precession by a temperature gradient would realize a quantum heat engine. Conversely, forcing a magnetic precession can be used to pump heat against a temperature gradient. Setups with several magnetic islands could be engineered to effect exchange of charge and energy without employing a {\em dc} battery. These devices have been explored in the literature on quantum pumps \cite{arr07,janine,mos09} and their efficiencies could be analyzed in the thermoelectric framework of Ref.\ \cite{recentus}. 

\section*{Acknowledgement}

We thank Gil Refael and Ari Turner for discussions. This work was supported by CONICET, MINCyT and UBACyT (L.A.) as well as the Deutsche Forschungsgemeinschaft and the Helmholtz Virtual Institute {\em New States of Matter and Their Excitations} (F.v.O.). L.A.\ thanks the ICTP Trieste for hospitality and the Simons Foundation for support. F.v.O.\ thanks the KITP Santa Barbara for hospitality during the final preparation of this manuscript. This research was supported in part by the National Science Foundation under Grant No.\ NSF PHY11-25915.

\end{document}